\documentclass{aastex}          
\usepackage{spr-astr-addons}    
\usepackage{url}

\begin{document}
%
\title{On the viability of a certain vector-tensor theory of gravitation}

\shorttitle{<vector-tensor tensor of gravitation>}
\shortauthors{R. Dale and D. S\'aez}

\author{Roberto Dale\altaffilmark{1}}      
\affil{Departamento de F\'{\i}sica, Ingenier\'{\i}a de Sistemas 
y Teor\'{\i}a de la Se\~nal, Universidad de Alicante, 03690, San Vicente del Raspeig, 
Alicante, Spain.}
\and 
\author{Diego S\'aez\altaffilmark{2}}
\affil{Departamento de Astronom\'\i a y Astrof\'\i sica, 
Universidad de Valencia, 46100-Burjassot, Valencia, 
Spain.}


\begin{abstract}
A certain vector-tensor theory is revisited.
Our attention is focused on cosmology. Against previous suggestions
based on preliminary studies, it is shown that, 
if the energy density of the vector field is large enough to
play the role of the dark energy and its fluctuations are negligible, 
the theory is not
simultaneously compatible with current observations on: supernovae, 
the cosmic microwave background (CMB) anisotropy, and the
power spectrum of the energy density fluctuations. 
However, for small enough energy densities 
of the vector field and no scalar fluctuations, 
the theory becomes compatible with all the  above observations
and, moreover, it leads to an interesting 
evolution of the so-called vector cosmological modes.
This evolution appears to be different from that of  
general relativity, and the difference might be useful to explain the
anomalies in the low order CMB multipoles.       
\end{abstract}


\section{Introduction}
\label{sec-1}
In previous papers \citep{mor07,mor08}, a number of effects  
produced by superhorizon cosmological vector modes were discussed 
in the framework of General Relativity (GR); in particular, 
it was proved that these modes may explain the anomalies 
in the first multipoles of the cosmic microwave background (CMB) 
temperature distribution. In GR, vector modes decay during the 
matter dominated era and, consequently, 
it is not easy to explain their presence 
at redshifts close to $z \simeq 1100$ 
(matter-radiation decoupling); however, 
in vector-tensor (VT) theories, the evolution of vector
modes is expected to be different and, in some cases, this evolution might 
be appropriate to explain CMB anomalies. It occurs in the case of
the VT theory studied in this paper; in fact, as it is proved 
below (see also \cite{dal09}), 
for appropriate small values of the vector field energy density
without scalar perturbations,
the theory is compatible with current observations and,
moreover, superhorizon vector modes
grow during the matter dominated era.

It is well known that the so-called {\em concordance model} 
simultaneously explains most of the current cosmological observations.
By this reason, this cosmological model is considered --along the paper-- 
as the preferred model
in the framework of GR. In the concordance model,  
the universe is quasi-flat and the initial fluctuations have an 
inflationary origin; in this situation, 
scalar perturbations are fully dominant, 
whereas the effects due to gravitational waves (vector modes) 
are expected to be small (negligible). Moreover, scalar perturbations are
purely adiabatic, their distribution is Gaussian, and  
the initial power spectrum
is similar to a Harrison-Zel'dovich one. 
There are a set of observations to be simultaneously explained 
by a concordance model; e.g.: (i) WMAP observations of the CMB
anisotropies \citep{kom10,jar10,lar10}, (ii) 
other CMB anisotropy observations --involving greater $\ell $-values-- 
performed from ground and balloon experiments; among them  
we mention  ACBAR \citep{rei08}, ACT \citep{fow10} and SPT \citep{lue09},  
(iii) high-redshift Ia supernovae (SNe Ia) observations \citep{ast06,rie06,kow08,fol09},
(iv) power spectra measurements based on galaxy surveys \citep{rei09},
(v) detection of baryonic acoustic oscillations (BAO) 
\citep{eis05,tak09,mar09,per09}, (vi) measurements of the Hubble constant
\citep{suy10}, and (vii) primordial deuterium abundance 
observations \citep{pet08}. In the LAMBDA archive (WMAP seven years data), various models 
--including different effect as lensing, Sunyaev-Zel'dovich, and so on--
are fitted to different sets of observational data including always the
proper WMAP7 data; e.g., for a $\Lambda $CDM model including lensing and 
Sunyaev-Zel'dovich effects, the best fitting to WMAP7 plus 
BAO and Hubble constant measurements corresponds to (see also \cite{jar10}):
(1) a reduced Hubble constant                                                  
$h=10^{-2}H_{0}=0.704^{+0.013}_{-0.014}$ (where $H_{0}$ is the                                         
Hubble constant in units of km s${}^{-1}$ Mpc${}^{-1}$); (2) 
density parameters $\Omega_{b} = 0.0456\pm 0.0016$, $\Omega_{c} =0.227 \pm 0.014$, and  
$\Omega_{\Lambda} = 0.728^{+0.015}_{-0.016}$ for the baryonic, dark matter, and
vacuum energy, respectively (the matter density parameter is then
$\Omega_{m} = \Omega_{b} + \Omega_{c} \simeq 0.2726$); (3) a total density parameter 
$\Omega_{_{T}} = 1.0023^{+0.0056}_{-0.0054}$,
(4) a parameter $\sigma_{8} = 0.809 \pm 0.024 $ normalizing 
the power spectrum of the scalar energy perturbations,
(5) a running scalar spectral index $n_{s} = 0.963 \pm 0.012 $ with 
$d n_{s} / d(ln \ k) = -0.022 \pm 0.020$ for $k=k_{0}=0.002 \ Mpc^{-1} $, 
(6) an equation of state for dark energy of the form $p = W \rho $ with
$W= -0.980 \pm 0.053$,
and (7) an optical depth  $\tau = 0.087 \pm 0.014$ 
which characterizes the reionization. 
There are fifty eight different fittings in the LAMBDA archive. All of them
give similar CMB angular power spectra to fit WMAP7 data. By this reason,
results in next sections appear to be almost independent of 
the particular choice 
of the fitting parameters, provided that they are
close enough to the central values of some LAMBDA fitting. 
Our choice of these parameters is done in section \ref{sec-3} and,
then, the resulting concordance model is assumed along the paper 
as a model of reference leading to a good fitting with current observations.
Predictions from models based on the VT theory under consideration
will be compared with observational data. Obviously, only predictions similar enough to 
those of the reference model may be compatible with current observations.

This paper is structured as follows: The VT theory is described in Section \ref{sec-2}, 
predictions of the concordance model are compared with observations in section \ref{sec-3}. 
These comparisons are 
extended to VT cosmology in section \ref{sec-4}. Vector perturbations are
studied in section \ref{sec-5} and, finally, section \ref{sec-6} is a general discussion 
and a summary of the main conclusions. 
Let us finish this section fixing some notation criteria. Latin (Greek) 
indexes run from 1 to 3 (0 to 3). The
gravitational constant and the scale factor are denoted $G$ and $a$, 
respectively. 
Units are chosen in such a way that the speed of light is $c=1$.

\section{The theory and its basic cosmological equations}   
\label{sec-2}

In vector-tensor theories,
there are two fields, the metric $g_{\mu \nu}$ and a four-vector $A^{\mu}$.
Various of these theories have been
proposed (see \cite{wil93}, \cite{wil06} and references cited there). We are
interested in one of the theories based on the action \citep{wil93}:
\begin{eqnarray}
I &=& ( {16\pi G} )^{ - 1} \int ( R + \omega A_\mu  A^\mu  R
+ \eta R_{\mu \nu }
A^\mu  A^\nu  -  \nonumber \\
& &
\varepsilon F_{\mu \nu } F^{\mu \nu }
+\gamma \,\nabla_\nu  A_\mu  \nabla^\nu  A^\mu
+ L_{m} ) \,\sqrt { - g} \,d^4 x
\label{1.1}
\end{eqnarray}  
where $\omega$, $\eta $, $\varepsilon$,
and $\gamma$ are arbitrary parameters, $R$,
$R_{\mu \nu}$, $g$, and $L_{m}$ are the scalar curvature, the Ricci tensor,
the determinant of the $g_{\mu \nu}$ matrix, and the matter Lagrangian, respectively.
The symbol $\nabla $ stands
for the covariant derivative, and
$F_{\mu \nu} = \nabla_{\mu} A_{\nu } - \nabla_{\nu} A_{\mu }$.
In action (\ref{1.1}), it is implicitly assumed that
the coupling between the matter fields and $A_{\mu}$ is negligible.
A different action is given in \cite{wil06}. It involves a new term of
the form $\lambda (A_{\mu}A^{\mu}+1)$, where $\lambda$ is a Lagrange multiplier.
With the help of this term, the vector $A^{\mu}$ is constrained to be
timelike with unit norm. From this last action,
the field equations of the so-called Einstein-Aether
constrained theories can be easily obtained. The
applications of these theories to Cosmology are discussed, e.g., in \cite{zun08}. A mass
term of the form $m_{_{A}}^{2} A_{\mu}A^{\mu} $ is used by \cite{boh07} to
explain cosmic acceleration with a massive vector field. The same acceleration 
was explained in the framework of the theory studied in this paper by
\cite{bel08}. Recently,
other theories involving vector fields have been also applied to cosmology
(see, e.g., \cite{tar07,mof06}).

Whatever the action parameters may be, only the PPN parameters 
$\gamma $, $\beta $, $\alpha_{1} $ and $\alpha_{2} $ may be 
different from those of GR in the
theories based on action (\ref{1.1}). Moreover, in these theories there is a varying
effective gravitational constant $G_{eff} $ (see \cite{wil93}).
On account of current solar system observations,
we are only interested in theories having $\gamma = \beta =1 $
and $G_{eff} = G$. This condition is hereafter 
our PPN viability condition. For  $\omega = \varepsilon = 0$,
there are only two theories compatible with the 
PPN viability condition, the first (second) of these theories 
corresponds to $\eta = - \gamma $ ($\eta =  \gamma $). 
In the second case, we have verified that vector cosmological 
modes evolve as in GR; hence, on account of our motivations 
(see section \ref{sec-1}),
we are not interested in this theory but in a theory with 
$\eta = - \gamma $. From the PPN parameters given in \cite{wil93},
it is easily seen that any theory with the action parameters
$\omega \neq 0$, arbitrary $\eta $, $\gamma = \eta $, and 
$\varepsilon = -\omega + \eta/2$ is compatible with 
the PPN viability condition. Finally, the same occurs for 
$\omega = 0$, arbitrary $\varepsilon $ and $\gamma = \eta $ and
also for $\omega = 0$, arbitrary $\varepsilon $, and
$\eta = 4\epsilon - \gamma$. Although all these theories deserve 
attention, their study is out of the scope of this paper. 
Here, we only analyze the theory with $\eta = - \gamma = 8 \pi G$
and $\omega = \varepsilon = 0$, whose field equations 
were given in \cite{dal09}. Various reasons have motivated
the choice of this theory: (1) it is the simplest theory
compatible with the PPN viability condition, (2)
it does not involve new dimensional constants, (3) 
it has been already studied in various papers \citep{bel08,bel09}, 
where it was proved that --in this theory-- there are no classical and 
quantum instabilities, and also that the so-called cosmic coincidence 
problem is avoided, and (4) deeper study is necessary 
to support its reliability. 

Scalar and vector $A^{\mu} $-perturbations of cosmological
interest should arise inside the effective horizon during inflation;
nevertheless, the VT theory does not define the inflationary period,
which is likely produced by some scalar field. The coupling between 
this scalar inflaton and the vector field $A^{\mu} $ is basic 
to estimate the initial inflationary $A^{\mu} $-perturbations.
This coupling is to be chosen among many possibilities.
After reheating, the scalar perturbations 
evolve coupled to the remaining scalar modes (associated to the metric, 
the energy densities, and so on) and their evolution is very complicated. 
A code similar to CAMB, but including the new scalar 
$A^{\mu}$-modes, would be necessary. This study and the full analysis 
of possible inflationary models is beyond 
the scope of this paper. On account of these facts, we first complete the study of
the selected VT theory in the case of negligible
$A^{\mu} $-perturbations (see previous papers by \cite{bel08} and \cite{bel09}) and, afterwards, 
we study the
evolution of possible vector perturbations after reheating. 
Encouraging results from this study (see sections \ref{sec-5} and 
\ref{sec-6}) suggest future work (see section \ref{sec-6}) in the framework of both 
the chosen VT theory and other VT theories 
compatible with the PPN viability condition.

From the field equations \citep{dal09}, one easily finds the equations describing
a homogeneous and isotropic universe, in which the line element is
\begin{equation}
ds^{2} = a^{2} \left[ -d\tau^{2} + \frac {dr^{2}} {1-Kr^{2}} +r^{2} d\theta^{2} 
+ r^{2} \sin^{2} \theta d\phi^{2} \right]
\label{bmetric} 
\end{equation}
and the vector field has the covariant components \\
$(A_{0}(\tau ), 0, 0, 0)$.
Constant $K$ takes on the values $+1$, $-1$ and $0$ in closed, open and flat 
universes, respectively, and $\tau $ is the conformal time.
The resulting equations are:
\begin{equation}
\ddot{A}_{0} = 4\, \frac {\ddot{a}}{a} \, A_{0}
\label{7.1}
\end{equation}
\begin{equation}
3 \frac {\dot{a}^{2}}{a^{2}} = 8\pi G a^{2} \left[ \rho_{_{F}} + 
\rho_{_{A}} + \rho_{_{K}}  \right]  
\label{7.2}
\end{equation}
\begin{equation}
-2 \frac {\ddot{a}}{a} +\frac {\dot{a}^{2}}{a^{2}} = 8\pi G a^{2} \left\{ p_{_{F}} 
+ p_{_{A}} + p_{_{K}} \right\} \ ,
\label{7.3}
\end{equation}  
where,
\begin{equation}
\rho_{_{A}} =
\frac {4 \dot{a}}{a^{5}}A_{0}\dot{A}_{0} - \frac {1} {2a^{4}}\dot{A}_{0}^{2}
-\frac{2 \dot{a}^{2}}{a^{6}}A_{0}^{2} \ , 
\label{7.4}
\end{equation}  
\begin{equation}
p_{_{A}} =
\frac {4 \dot{a}}{a^{5}}A_{0}\dot{A}_{0} 
- \frac {3} {2a^{4}}\dot{A}_{0}^{2}
-2 \left( 2\frac {\ddot{a}}{a^{5}} +  
\frac{ \dot{a}^{2}}{a^{6}}\right)A_{0}^{2} \ ,
\label{7.5}
\end{equation}   
\begin{equation}
\rho_{_{K}} =-3 p_{_{K}} = -3K/8 \pi G a^{2} \ ,
\label{7.6}
\end{equation}   
the dot stands for a derivative with respect to the time $\tau $,
and $\rho_{_{F}}$ and $p_{_{F}}$ are the energy density and pressure of the cosmological fluid,
respectively. This fluid contains baryons, radiation, and cold dark matter (CDM).
We see that the field equations of the theory couple the variables $A_{0}$ and $a$ 
describing the background. The density 
parameters due to baryons, radiation, CDM, the vector field and the curvature 
term are $\Omega_{b} $, $\Omega_{r} $,
$\Omega_{c} $, $\Omega_{_{A}} $, and $\Omega_{_{K}} $, respectively.
They satisfy the relation $\Omega_{b} + \Omega_{r}+\Omega_{c}+\Omega_{_{A}} 
+\Omega_{_{K}}= 1$. In the flat case one has $\Omega_{_{K}} =0$ and 
$\Omega = \Omega_{b} + \Omega_{r}+\Omega_{c}+\Omega_{_{A}}= 1$, 
whereas in the 
closed (open) case the curvature density parameter is negative (positive) and 
the inequality $\Omega = \Omega_{b} + \Omega_{r}+\Omega_{c}+\Omega_{_{A}}> 1$
($\Omega < 1$) is satisfied.

The free parameters of the theory are assumed to be the Hubble constant and the 
density parameters $\Omega_{_{A}}$ and $\Omega_{_{K}}$.  
In the flat case, the present value of the scale factor, $a_{0}$, is 
arbitrary (we assume $a_{0}=1$); whereas in nonflat models this value is 
$a_{0} = H_{0}^{-1} |\Omega_{_{K}}|^{-1/2}$.
For a given choice of the free parameters, Eqs. (\ref{7.1}) -- (\ref{7.3}) may be
numerically solved to get functions $a=a(\tau )$ and $A_{0} = A_{0}(\tau )$.
Finally, by using Eqs. (\ref{7.4}) -- (\ref{7.5}), the equation of state for
the dark $A^{\mu}$-energy; namely, the ratio
$W(\tau) = p_{_{A}}(\tau) / \rho_{_{A}} (\tau)$ may be easily calculated.
We can then state that, in the absence of $A^{\mu} $-perturbations,
the VT theory is cosmologically equivalent 
to GR plus dark energy with the computed $W(\tau)$ 
ratio. This variable $W(\tau )$ may be easily used 
for numerical calculations with CAMB code \citep{lew00}.
This code allows us to calculate all the cosmological spectra, 
including those associated to the 
CMB polarization. All these spectra are influenced by 
the vector field of the VT theory. The effects of the vector field 
would be identical to those produced, in GR, 
by dark energy with the associated equation of state 
($W(\tau )$ ratio computed in the VT theory).

\section{Supernovae, CMB and Power Spectrum Observations}
\label{sec-3}

The concordance model is now defined and analyzed in detail. The 
contents of this section are used along the paper to discuss the 
viability of the vector-tensor theory under consideration. Any admissible theory 
must explain the same observations as  
the concordance model with comparable accuracy. Hereafter, this model is 
defined by the following 
set of parameters, which are close to the central values for the fitting 
considered in section II: h=0.704, 
$\Omega_{b} = 0.0461$, $\Omega_{c}= 0.2265$, $\Omega_{k}= -0.0012$, 
$\tau = 0.087 $, $W= -1$, $n_{s} = 0.96 $,
$d n_{s} / d(ln \ k) = -0.016 $ and $\sigma_{8} = 0.8065 $. 
These parameters are used to get the angular correlations 
of the CMB and the power spectrum $P(k) $ of the matter fluctuations.
Calculations are performed by using CAMB (see section \ref{sec-2}). 
The relation between the distance modulus $\mu = m-M$ and 
the redshift $z$ of the $Ia $ supernovae only depends on the 
parameters $h$, $\Omega_{m} $ and $\Omega_{k}$. 
In other words, this relation only depends on some parameters describing the
background universe, whereas the parameters associated to 
perturbations are irrelevant. Let us now study the agreement of this 
model with data from various observations.

\begin{table*}
\small
\caption{Parameter configurations in GR and VT cosmologies.\label{tab:table1}}
\begin{tabular}{@{}crrrrrrrrrrrr@{}}
\tableline
Model&$\Omega_{m}$&$h$&$\Omega_{b}h^2$
&$\Omega_{c}h^2$&$\Omega_{_{\Lambda}}$&$\Omega_{_{A}}$&$\tau$&$\sigma_8$&
$n_{s}$&$d n_{s} / d(ln \ k)$&$\Omega_{k}$ & $\chi_{_{CMB}}^{2}$ \\   
\tableline
GR1&.2726&.704&.0228&.1123 
&.7286& &.087 & .8065 & .96 & -.016 & -.0012 & 36.81\\ 
VT1& .25 & .82 & .0223 & .1458 & 0. & .7512 & .087 & .8065 & .96 & -.016 & -.0012 & 15917.\\ 
VT2& .25 & .82 & .0223 & .1458 & 0. & .7512 & .000 & .95 & .96 & -.016 & -.0012 & 2467. \\ 
VT3& .25 & .82 & .0223 & .1458 & 0. & .7512 & .000 & .95 & .92 & -.016 & -.0012 & 1473.\\ 
VT4& .25 & .82 & .0223 & .1458 & 0. & .7512 & .000 & .95 & .84 & -.016 & -.0012 & 432.8\\ 
VT5& .25 & .82 & .0223 & .1458 & 0. & .7512 & .000 & .95 & .84 & -.05 & -.0012 & 288.4 \\
VT6& .25 & .82 & .0223 & .1458 & 0. & .7512 & .000 & .95 & .84 & -.09 & -.0012 & 180.9\\
VT7& .25 & .82 & .0223 & .1458 & 0. & .7700 & .000 & .95 & .84 & -.09 & -.02 & 1057.9\\
VT8& .25 & .82 & .0223 & .1458 & 0. & .7300 & .000 & .95 & .84 & -.09 & .02 & 355.0\\
VT9& .25 & .82 & .0223 & .1458 & 0. & .7400 & .000 & .942 & .84 & -.09 & .01 & 83.28\\
VT10& .36 & .73 & .0223 & .1695 &0.  & .6412 & .087 & .8065 & .96 & -.016 & -.0012 & 17725. \\ 
VT11& .36 & .73 & .0223 & .1695 &0.  & .6412 & .000 & .95 & .96 & -.016 & -.0012 & 3802. \\ 
VT12& .36 & .73 & .0223 & .1695 &0.  & .6150 & .000 & .963 & .84 & -.13 & .025 & 86.62\\
VT13& .48 & .67 & .0223 & .1932 &0.  & .5212 & .087 & .8065 & .96 & -.016 & -.0012 & 19683.\\
VT14& .48 & .67 & .0223 & .1932 &0.  & .5212 & .000 & .95 & .96 & -.016 & -.0012 & 5412. \\
VT15& .48 & .67 & .0223 & .1932 &0.  & .4800 & .000 & .993 & .84 & -.16 & .04 & 136.9\\
VT16& .2726 &.704 &.0228 &.1123 &.7276 & .0010 &.087 & .8065 & .96 & -.016 & -.0012 & --- \\ 
\tableline
\end{tabular}
\end{table*}

\begin{figure*}[tb]
\begin{center}
\resizebox{0.6\textwidth}{!}{%
  \includegraphics{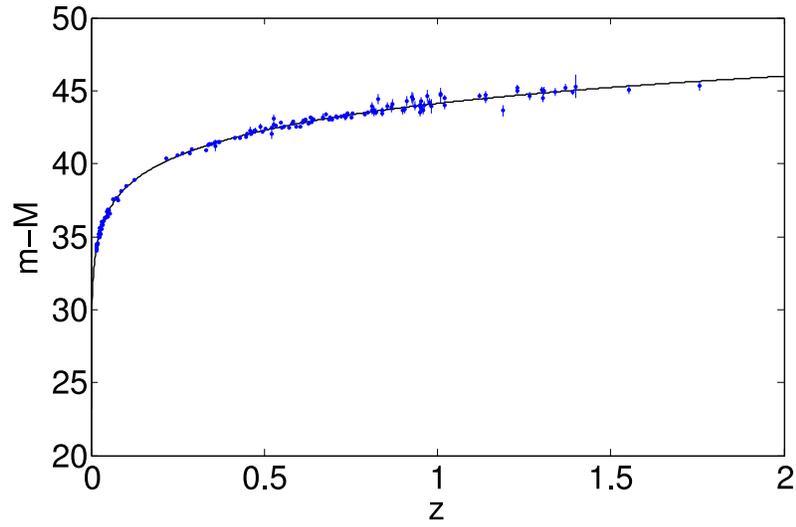}
  } 
\caption{ Solid line:
supernova distance modulus, $\mu$, 
in terms of the redshift, z, for the concordance model. Observational data
from the supernova legacy survey \citep{ast06} and from a high redshift Hubble space 
telescope sample \citep{rie06} are also shown.
}
\label{figu1}
\end{center}
\end{figure*}

For the concordance model (h=0.704, $\Omega_{m} = 0.2726$ and $\Omega_{k}= -0.0012$), 
the relation
$\mu = \mu (z) $ is represented in the solid line of Fig. \ref{figu1}. In the same Figure
we exhibit 156 observational data, from \cite{ast06} and \cite{rie06}, which
cover a wide range of redshifts from $z=0.015$ to $z=1.755$. 
Error bars do not include the 
uncertainty in the absolute magnitude of the SNe Ia, which is typically
$0.15 \ mag$, we then calculate the following function (see \cite{ast06})
\begin{equation}
\chi_{_{SN}}^{2} = \sum_{_{SNe}} \frac {(\mu_{obs} - \mu_{the})^{2}} {\sigma^{2}(\mu) + \sigma^{2}_{int} } \ ,
\end{equation} 
where $\mu_{obs}$ ($\mu_{the}$) is the observational (theoretical) $\mu $ value, $\sigma_{int} $
accounts for the intrinsic dispersion of SNe Ia and $\sigma(\mu)$ is the error represented
in Fig. \ref{figu1}. For $\sigma_{int} = 0.15 \ mag$, the 
$\chi_{_{SN}}^{2} $ value is $132.45$ and the $p$-value is $0.91 $. 
This is a large value suggesting that the chosen parameters explain 
very well the SNe Ia observations. Of course, these parameters are 
very similar to those of section \ref{sec-1}, which are compatible 
--by construction-- with WMAP7, BAO, and $H_{0}$ observations.

\begin{figure*}[tb]
\begin{center}
\resizebox{0.6\textwidth}{!}{%
  \includegraphics{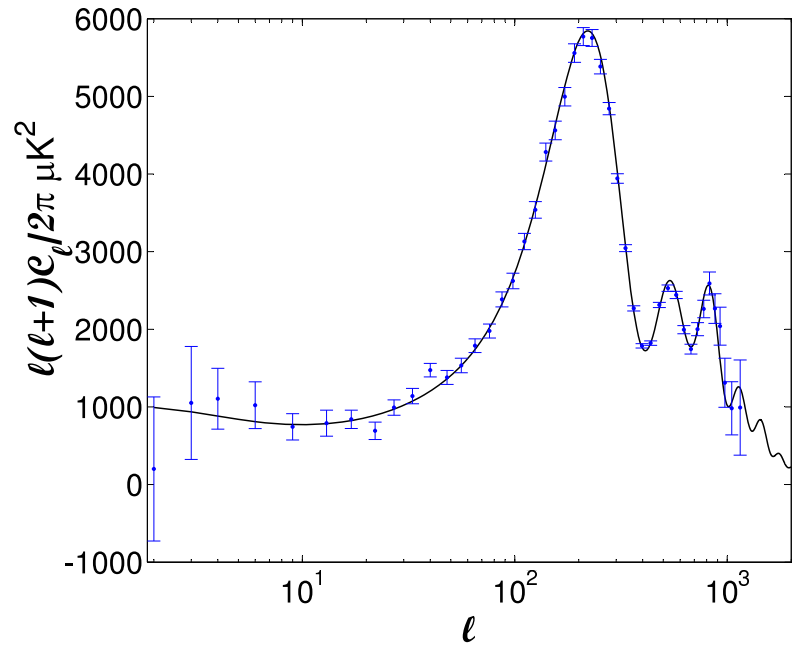}
  } 
\caption{Solid line: CMB angular power spectrum 
for the concordance model including lensing. Observational data (WMAP seven years)
and error bars (including cosmic variance) are also shown. 
}
\label{figu2}
\end{center}
\end{figure*}

Let us now study the CMB angular power spectrum for the 
concordance model. CAMB calculations are performed including lensing.
The resulting spectrum is given in the solid line of 
Fig. \ref{figu2}, where we also present the binned WMAP seven years measurements 
with the corresponding error bars, which can be found in LAMBDA 
(Legacy Archive for Microwave Background Data Analysis). 
The comparison between observations and theoretical predictions is 
performed by using the function
$\chi_{_{CMB}}^{2} = \sum_{bins} (\Delta_{obs} - \Delta_{the})^{2}/ \sigma^{2}$, 
where $\Delta = \ell (\ell +1) C_{\ell} /2 \pi$ and
$\Delta_{obs}$ ($\Delta_{the}$) stands for the observational (theoretical)
$\Delta$-values, respectively.
The dispersion of $\Delta $ inside each $\ell$-bin is 
measured by $\sigma^{2}$. From the 45 bins 
considered by the WMAP team, the resulting $\chi_{_{CMB}}^{2} $ is 36.81 and the 
corresponding p-value is $0.80$. This high p-value strongly suggest that
the concordance model fits very well WMAP7 observations.
                                           
The spectra corresponding to the electric part of the 
CMB polarization and the cross correlation $TE$ have been also obtained, but 
they are not necessary for the discussion presented in this paper. 

\begin{figure*}[tb]
\begin{center}
\resizebox{0.8\textwidth}{!}{%
  \includegraphics{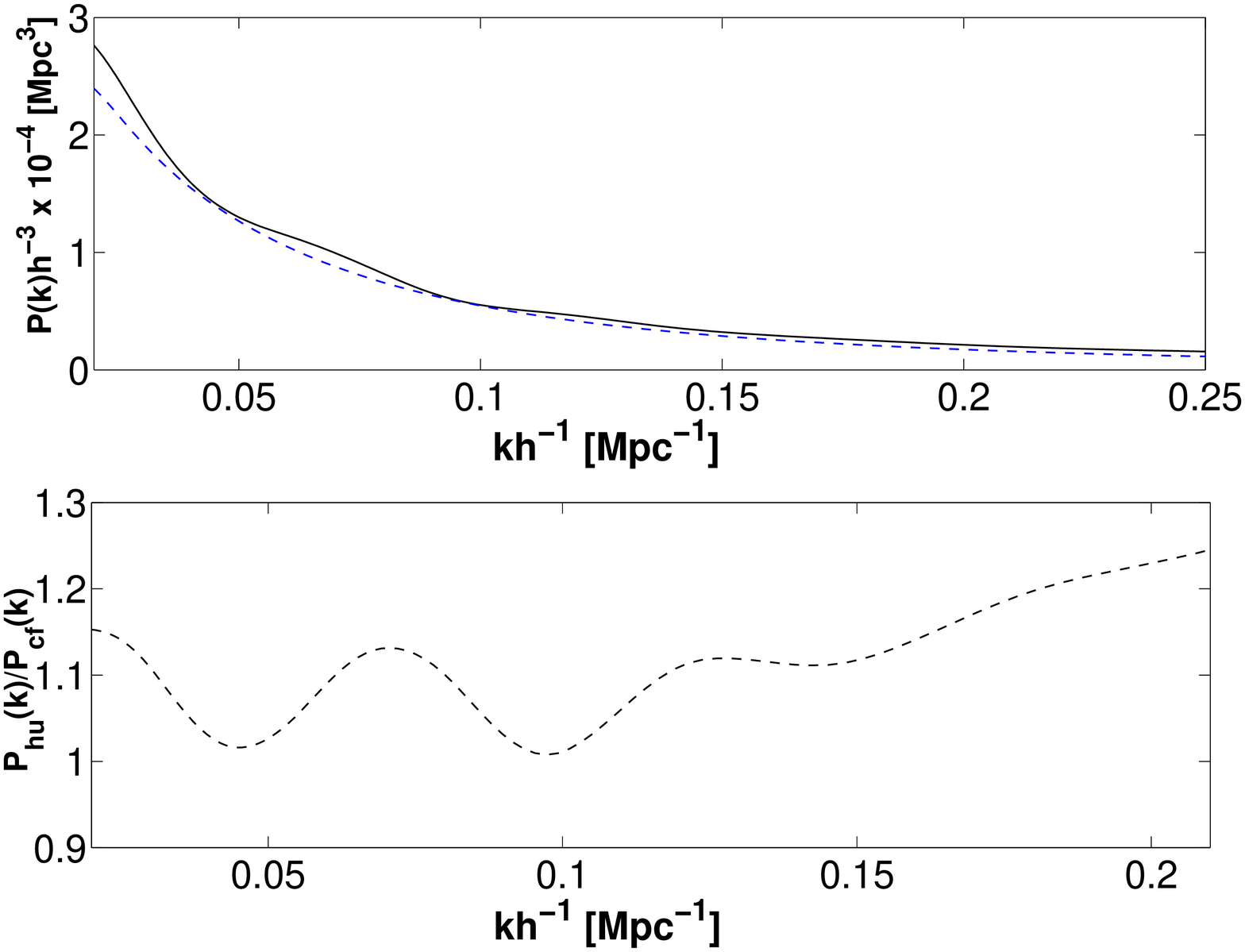}
  } 
\caption{Top: Power spectrum of the matter energy density
perturbations in the concordance model. Solid 
line is the true power spectrum estimated with CAMB. Dashed line 
shows the power spectrum --for the same model-- 
in the absence of baryon acoustic oscillations.
Bottom: Ratio between the power spectra showed in the to panel  
}
\label{figu4}
\end{center}
\end{figure*}

The matter power spectrum estimated with CAMB  
is shown in the solid line of Fig. \ref{figu3} together with
observational data \citep{eis05} from the 
sloan digital sky survey (SDSS). 
Baryonic acoustic 
oscillations (BAOs) would produce the deviation between the solid (CAMB spectrum)
and the dashed  
lines of the top panel of Fig. \ref{figu4}. The dashed line shows the 
spectrum in the absence of baryon acoustic oscillations \citep{hu98,ei98}. 
Particularly relevant is the deviation appearing 
in the $k/h$ interval (0.05, 0.1). 
Recent observations seems to support the existence 
of this feature \citep{eis05,mar09,per09}, which has been taken into account 
(by the WMAP team) to 
find the parameters of section \ref{sec-1}.

As it is shown in Fig. \ref{figu3}, the observed and theoretical power
spectra fit very well for large spatial scales with $k < 0.05 h \ Mpc^{-1}$,
which are clearly linear. For smaller scales,
the error bars of the observational data are above the solid line (predictions 
based on the concordance model). The theoretical spectrum 
is due to perturbations in both dark and baryonic matter,
whereas the power observed in galaxy surveys corresponds 
to fluctuations in the baryonic component alone.
These two spectra would be only identical in the absence of any bias between 
dark and baryonic energy densities at the corresponding spatial scales. 
Hence, if there are no unknown systematic errors in 
the observations, Fig. \ref{figu3} suggests a small bias, for
$k > 0.05 h \ Mpc^{-1}$,
whose origin has not been explained: We might speculate with some kind of 
nonlinear effect, with a primordial bias depending on the spatial scales
(which vanishes for $k < 0.05 h \ Mpc^{-1}$), 
and so on.

Another characteristic of the matter power spectrum is the $\sigma_{8} $ value,
which is essentially related to the power at scales smaller 
than $8 h^{-1} \ Mpc$; namely, to the 
power corresponding to wavenumbers $k> 0.125h \ Mpc$. 
Various estimates of $\sigma_{8} $ have been reported in the technical 
literature.
After seven years of CMB observations, WMAP team has obtained the value
$\sigma_{8} \simeq 0.8$. Moreover, recent measurements of the CMB 
angular power spectrum, at very small angular scales \citep{fow10,lue09}, suggest
$\sigma_{8} $ values smaller than $\sim 0.86 $. Similar bounds have been 
obtained from the analysis of ROSAT X-ray cluster data (see \cite{fow10}
and references cited therein). Finally, the $\sigma_{8} $ value corresponding to the 
SDSS data of Fig. \ref{figu3} is close to 0.84.
On account of all these results,  
we hereafter accept the observational constraint
$\sigma_{8} < 0.9$ for dark plus baryonic matter.

For a given theoretical model, 
the resulting p-values depend on the chosen observational data,
but the data used along this section have 
been selected among the most accurate current observations
(WMAP for the CMB, SDSS for the galaxy distribution, and 
the supernova legacy survey and Hubble observations for supernovae) and, 
consequently, we can be 
confident with our results and use the same data to study 
the compatibility between theoretical predictions and observations  
in the framework of the VT theory under consideration. It is done in
next section.

\section{Results}
\label{sec-4} 

We begin with the study of the apparent supernovae dimming in the
framework of the VT theory described in section \ref{sec-2}. 
In this theory and also in any theory 
with cosmological models based on the Robertson-Walker background metric,
CMB observations ensure that the universe is almost flat. It is due to the 
fact that the first peak of the CMB angular power spectrum appears located at  
$\ell \simeq 210$. On account of these facts, 
the value of the curvature density parameter
has been chosen to be $\Omega_{k} = -0.0012 $ as in our version of the 
concordance model. In the VT theory, CAMB numerical calculations 
can be easily performed in the presence of this
small curvature and, consequently, it has been 
maintained all along this section. Nevertheless, it has been verified 
that, with the chosen $\Omega_{k}$-value, results are almost indistinguishable from
those corresponding to a strictly flat universe (see first paragraph of section \ref{sec-5}). 
Parameters $h$ and $\Omega_{m} $ have been 
varied. Only parameters $\Omega_{k}$, $h$, and $\Omega_{m}$
are relevant in supernova studies.

\begin{figure*}[tb]
\begin{center}
\resizebox{0.6\textwidth}{!}{%
  \includegraphics{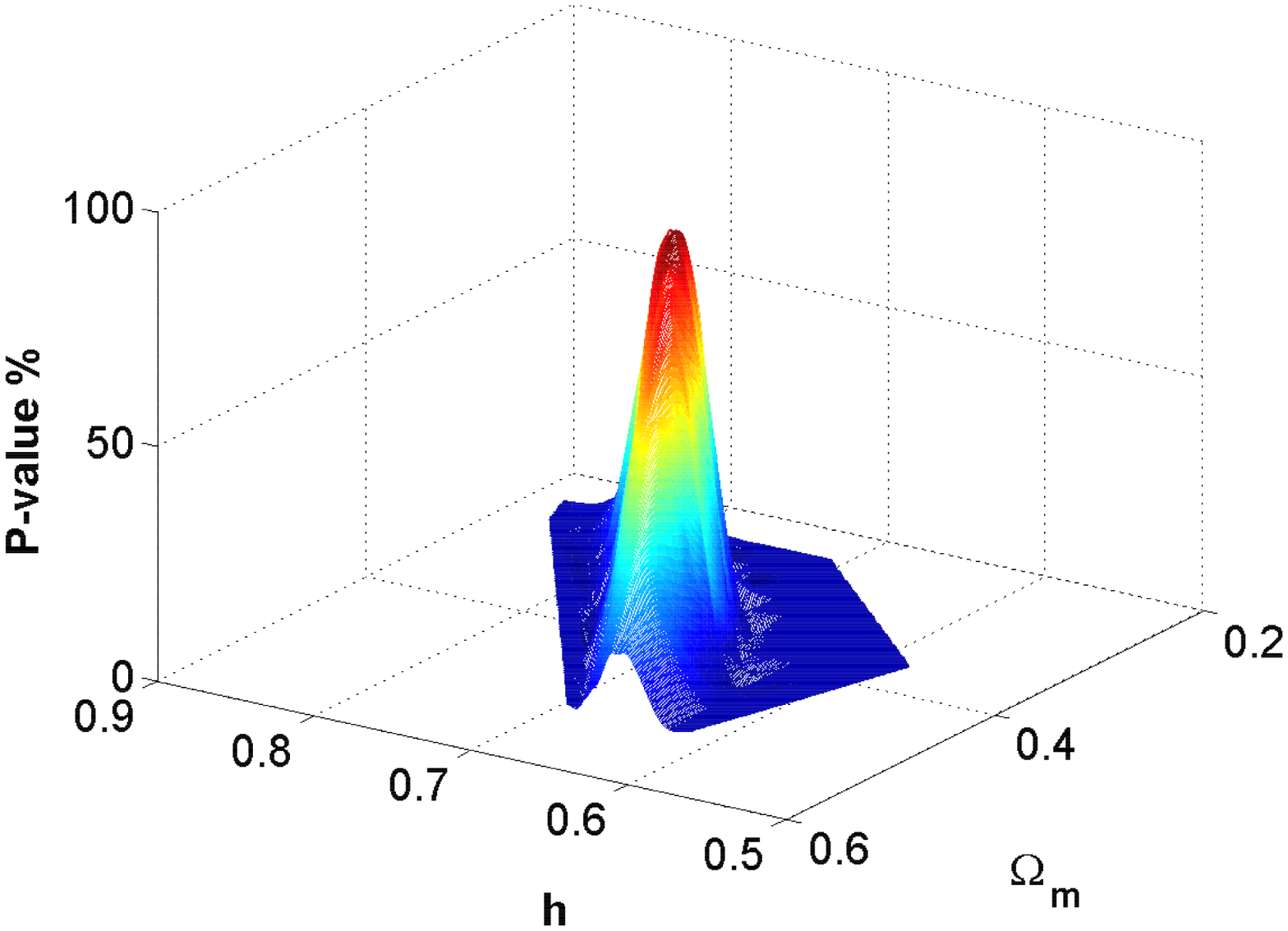}
  }
\caption{3D representation of quantity $100p$ 
{\it v.s.} parameters $\Omega_{m}$ and $h$ for Ia SNe. 
Gravitation is described by the vector tensor theory of 
section \ref{sec-2}}
\label{figu5}       
\end{center}
\end{figure*}

\begin{figure*}[tb]
\begin{center}
\resizebox{0.6\textwidth}{!}{%
  \includegraphics{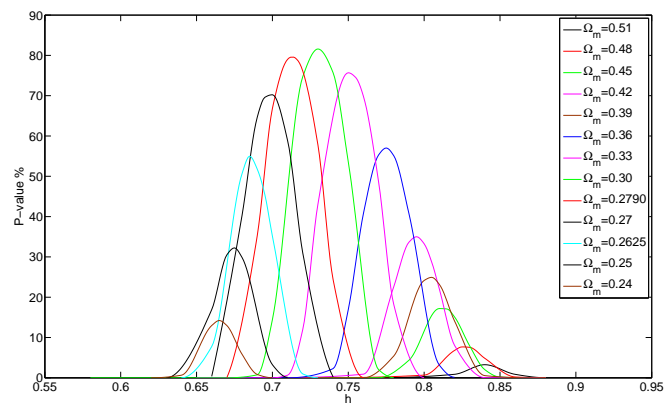}
  }
\caption{Quantity $100p$ {\it v.s.} $h$ for 
various $\Omega_{m}$ values. Each curve corresponds to an 
$\Omega_{m}$ value as it is displayed inside the panel. }
\label{figu7}       
\end{center}
\end{figure*}

For each pair ($h$, $\Omega_{m} $), function $\chi_{_{CMB}}^{2} $ and the associated p-value 
have been calculated. The same observational data as in section \ref{sec-3} have been 
used in the computations. 
From a purely statistical point of view,
models leading to $p< 0.05$ are usually ruled out.
In Fig. \ref{figu5} you can see a 3D representation 
involving quantities $h$, $\Omega_{m}$, and  $100p$ (p-value \%). In 
this Figure, we see that the p-value is larger than $0.05$ in a bounded region 
of the ($h$, $\Omega_{m} $) plane. This region is 
exhibited in Fig. \ref{figu6}, where the $100p$-values in the
($h$, $\Omega_{m} $) plane are represented by using a color scale.
For any small $\Omega_{k}$ parameter allowed by
CMB observations,
the shape and size of this region are very similar.
According to Fig. \ref{figu7}, condition
$p< 0.05$ rules out $\Omega_{m} $ values smaller than $\sim 0.24$. We also see
that $\Omega_{m}$-values a little greater than $0.48$ are
not discarded by the same condition; however, 
they will be discarded from our analysis of the 
CMB angular power spectrum (see section \ref{sec-4}).

Let us now study the CMB angular power spectrum for 
all the pairs ($h$, $\Omega_{m} $) explaining supernovae observations. 
We have studied forthy two ($\Omega_{m},h$) pairs corresponding to seven 
h-values and six $\Omega_{m}$-values
uniformly spaced in the intervals [$0.6,0.9$], and  [$0.25,0.55$], respectively.
These pairs cover the region of the $h$-$\Omega_{m} $ plane 
allowed by SNIa observations for any small 
$\Omega_{k}$-value compatible with the first peak location in the 
CMB angular power spectrum. We cover a region a little greater than 
that described in Figs. \ref{figu5}, \ref{figu6}, and \ref{figu7}.
Each ($\Omega_{m},h$) pair is studied by using the same method, which 
is now described in detail for three selected additional pairs 
located in the region of Fig. \ref{figu6}:
In the top left panel of Fig. \ref{figu8}, we show $C_{\ell} $ quantities for the 
pair ($h= 0.82 $, $\Omega_{m}= 0.25 $), which poorly explains supernovae observations with
$p \simeq 0.07$. 
The middle left panel corresponds to $h= 0.73 $ and $\Omega_{m}= 0.36 $ 
and supernova observations are well
explained with $p \simeq 0.82 $  and, finally, in the bottom left panel we have 
represented the case $h= 0.67 $ and $\Omega_{m}= 0.48 $ 
having $p \simeq 0.31$. These three pairs correspond to the maxima of the 
curves presented in Fig. \ref{figu7} for $\Omega_{m}= 0.25 $, $\Omega_{m}= 0.36 $,
and $\Omega_{m}= 0.48 $.

Let us begin with the pair $h= 0.82 $, $\Omega_{m}= 0.25 $, which is 
not a good choice from the point of view of supernovae and $h$ measurements 
(see \cite{suy10}); however, it appears to be the best choice to deal with
the CMB (see below). By this reason, this pair plays a central role in 
our discussion on CMB anisotropy. The dotted line of the top left panel 
of Fig. \ref{figu8} is the angular power spectrum for the
parameter configuration VT1 of Table \ref{tab:table1}. In this configuration, 
the value of $\Omega_{b} h^{2} $ is assumed to be $0.0223 $, which is 
the maximum possible value at 68 \% confidence level according to \cite{pet08},
the $\Omega_{c} $ value is then fixed by 
the relation $\Omega_{m} = \Omega_{b} + \Omega_{c} =0.25$ and, 
the remaining 
parameters are taken identical to those of the concordance model.
In the same panel, and also in any left panel of Figs. \ref{figu8}--\ref{figu10}, 
the solid line shows the angular power spectrum 
of the concordance model (GR1 entry of Table \ref{tab:table1}). 
The same is valid for any right panel, where the solid line is
the matter power spectrum of the case GR1.
We easily see that the chosen parameters lead to very small 
$C_{\ell} $ quantities (dotted line) which cannot explain WMAP observations.
Accordingly, for                                                            
these parameters and the observational data of Fig. \ref{figu2}, we have obtained 
$\chi_{_{CMB}}^{2} = 15917. $ and $p \simeq 0.$ \, . 
In configurations VT1-VT15
of Table \ref{tab:table1}, the parameter $\Omega_{b} h^{2} $ takes
on its maximum possible value \cite{pet08}. Other possible values 
are considered in the last paragraph of this section.

We can now modify some of the remaining parameters
to look for a good set of $C_{\ell} $ coefficients. Actually, the following 
parameters may be changed: $\tau $, $\sigma_{8} $, $\Omega_{k}$, $n_{s} $, and                        
$d n_{s} / d(ln \ k) $. Changes in the parameters  
$\tau $ and $\sigma_{8} $ may produce a large magnification 
of the angular
power spectrum amplitude. Since this amplitude increases as $\sigma_{8} $ ($\tau$)
increases (decreases), the values $\tau = 0$ and $\sigma_{8}=0.95$ lead 
to a relevant magnification, which is not realistic as a result of
the following facts: (i) the inequality 
$\sigma_{8}<0.9$ should be satisfied
(see section \ref{sec-3}) and, (ii) reionization at a redshift $z \simeq 6$ 
is proved by measurements of the Gunn-peterson effect \citep{bec01}, which
implies $\tau > 0$.
For the chosen new 
extreme values of $\tau $ and $\sigma_{8} $, plus 
the values $n_{s} = 0.96$ and                         
$d n_{s} / d(ln \ k) = -0.0016 $ of the concordance model
(parameter configuration VT2 of Table \ref{tab:table1}), 
the $C_{\ell} $ quantities have been calculated, they 
are shown 
in the dashed line of the top left panel of Fig. \ref{figu8}.
We see that, in spite of our forced choice of
$\tau $ and $\sigma_{8} $,  
the resulting angular power spectrum remains 
smaller than that of the concordance model (solid line).
In this case, we have found $\chi_{_{CMB}}^{2} = 2467. $ and $p \simeq 0$.

The same study has been performed for the pairs 
($h= 0.73 $, $\Omega_{m}= 0.36 $) and ($h= 0.67 $, $\Omega_{m}= 0.48 $).
Results are presented in the left middle and left bottom panels 
of Fig. \ref{figu8}. In the middle left (bottom left) panel the dotted line
is found for the configurations VT10 (VT13) of Table \ref{tab:table1},
whereas, the dashed lines are obtained for the 
configuration VT11 (VT14). As in the top left panel, the dashed lines
correspond to $\tau = 0$ and $\sigma_{8}=0.95$. 
For these lines, we have found ($\chi_{_{CMB}}^{2} = 3802. $, $p \simeq 0.$) and
($\chi_{_{CMB}}^{2} = 5412. $, $p \simeq 0.$) for the middle left
($h= 0.73 $, $\Omega_{m}= 0.36 $)
and the bottom left ($h= 0.67 $, $\Omega_{m}= 0.48 $)
panels, respectively. These $\chi_{_{CMB}}^{2} $-values are larger than 
that corresponding to $h= 0.82 $ and $\Omega_{m}= 0.25 $ ($\chi_{_{CMB}}^{2} = 2467. $);
hence, the best situation is found for this last pair. 
The same occurs for any pair compatible with SNe Ia observations.

After these considerations we continue with the study of the
best pair. The question is: Could we fit the observations 
varying  $\Omega_{k}$, $n_{s} $, and                         
$d n_{s} / d(ln \ k) $ in the parameter configuration VT2? 
We begin with parameter $n_{s} $, whose value, in VT2, is
$n_{s}=0.96 $. In the new 
configurations VT3 and VT4 we have taken $n_{s}=0.92 $ and
$n_{s}=0.84 $, respectively. The remaining parameters 
have not been altered (see Table \ref{tab:table1}).
The resulting angular power spectra are exhibited in 
the top left panel of Fig. \ref{figu9}. 
The dashed (dotted) line corresponds to $n_{s}=0.92 $ ($n_{s}=0.84 $). 
For the VT3 (VT4) model we have found $\chi_{_{CMB}}^{2} = 1473. $ ($\chi_{_{CMB}}^{2} = 432.8 $) and
$p \simeq 0.$ ($p \simeq 0.$). Accordingly, we see that the dashed line is well below the solid 
line. The dotted line approaches rather well the first peak,
but it does not fit the low $\ell $ multipoles and the remaining 
peaks of the solid line. A good fitting  based 
on parameter $n_{s} $ is not possible for any $\ell < 1150 $ 
(the largest $\ell $ value in the WMAP seven years binned data).

It is also remarkable that the value $n_{s} = 0.84 $, which leads to a good
fit of the first peak, is too small 
to be compatible with standard inflationary models, for which, 
$n_{s} > 0.94$ \citep{pa07}. Rather exotic inflationary models \citep{mu06}, whose 
Lagrangians involve nonlinear functions of 
the inflaton kinetic energy might be a rare exception.

Let us now change the value of $d n_{s} / d(ln \ k) $ in 
configuration VT4 of Table \ref{tab:table1}. Two new values of
$d n_{s} / d(ln \ k)$: $-0.05 $ and $-0.09 $ have been assumed in 
configuration VT5 and VT6, respectively. The angular power spectrum of these two new configurations 
are shown in the left middle panel of Fig. \ref{figu9}. The dashed (dotted-dashed) line
corresponds to the value $-0.05 $ ($-0.09 $). The dotted line is identical to 
the corresponding line of the top left panel.
Comparison with WMAP binned data gives $\chi_{_{CMB}}^{2} = 288.4$ ($\chi_{_{CMB}}^{2} = 180.9$) for the value 
$-0.05 $ ($ -0.09$) and $p \simeq 0.$ in both cases.

As a last step, the parameter $\Omega_{k} $ has been slightly changed in configuration VT6.
The new values -0.02 (closed universe) and 0.02 (open universe) have been 
assumed in the configurations VT7 and VT8 of Table \ref{tab:table1}, respectively.
Results are presented in the bottom left panel of Fig. \ref{figu9}. 
The dotted (dashed) line corresponds to the closed (open) case. In the closed (open)
case, the peaks are lower (higher) than in the concordance model and they are 
shifted to left (right). The low $\ell $ multipoles are almost independent of
these changes in $\Omega_{k}$. Since too large peak shifts are inadmissible, parameter
$\Omega_{k} $ is constrained to be smaller than a few hundredths.

Finally, we have varied parameters $\sigma_8 $ and $\Omega_{k} $ in configuration 
VT8 to optimize the fitting to the observational data. The resulting values 
$\sigma_8 = 0.942$ and $\Omega_{k} = 0.01 $ (slightly open case) define the 
VT9 configuration. The angular power spectrum of this last configuration is 
displayed in the dotted line of the top left panel of Fig. \ref{figu10}.  
In cases ($h= 0.73 $, $\Omega_{m}= 0.36 $) and ($h= 0.67 $, $\Omega_{m}= 0.48 $),
the best fittings have been found for ($\sigma_8 = 0.963$, $\Omega_{k} = 0.025 $)
and ($\sigma_8 = 0.993$, $\Omega_{k} = 0.04 $), respectively. The resulting 
spectra  
are displayed in the dotted line of the middle left ($h= 0.73 $, $\Omega_{m}= 0.36 $) 
and bottom left ($h= 0.67 $, $\Omega_{m}= 0.48 $)
panels of Fig. \ref{figu10}. In the top left panel ($h= 0.82 $, $\Omega_{m}= 0.25 $), 
we see a rather good fitting to the observational data excepting the first ten 
multipoles. As a result of this discrepancy one has $\chi_{_{CMB}}^{2} = 83.28$ and 
$p \simeq 4.52 \times 10^{-4} $. In the middle left and bottom left panels 
the fittings are slightly worse in the regions of both 
the first multipoles and the third peak. Accordingly, we have 
found $\chi_{_{CMB}}^{2} = 86.62$ for the middle left panel and $\chi_{_{CMB}}^{2} = 136.9$
for the bottom left one. All these $\chi_{_{CMB}}^{2} $-values are much greater than $36.81$,
which is the value corresponding to the concordance model. For pairs ($h$, $ \Omega_{m} $)  
with $\Omega_{m}> 0.48 $, the fittings are worse.

The matter power spectra of all the configurations considered in the above discussion 
have been presented, with the same type of line, in the corresponding right panels
of Figs. \ref{figu8}, \ref{figu9}, and \ref{figu10}.
From these figures, it follows that the matter power spectrum and the CMB angular 
power spectrum do not tend
to the spectra of the concordance model at the same time. The VT configurations considered 
in the dotted lines of Fig. \ref{figu10} lead to the best fittings to the CMB multipoles, but
the matter powers of the same configurations (dotted line of the right panels)
appear to be too large for any $k$-value.
For large spatial scales $k<0.05h \ Mpc^{-1}$, the observational data fit very well the 
predictions of the concordance model (solid lines); however, 
the predictions of the VT theory (dotted lines)
are fully incompatible with the same data.
The situation is much worse than in the concordance model.

Best results have been obtained for 
$\sigma_{8} > 0.9$, $\tau =0$, and too large (small) values of $n_{s}$ 
($d n_{s} / d(ln \ k) $); namely, for unrealistic values of the above 
four parameters. Let us now consider new values of 
$\Omega_{b} h^{2} $, which must be smaller than 0.0223.
We have verified that (as it was expected) the  $C_{\ell} $ coefficients of 
three new configurations similar to VT1, VT10 and VT13 but having 
$\Omega_{b} h^{2} <0.0223$ 
are smaller than those of the proper VT1, VT10 and VT13
configurations. Then, starting from the new smaller coefficients,
the same process --used above to look for the best fittings to
the CMB and $P(k)$ observed spectra-- leads to  
$\sigma_{8} $, $\tau $, $n_{s}$ and $d n_{s} / d(ln \ k) $ values
less realistic than in the case $\Omega_{b} h^{2} =0.0223$; hence,
we cannot explain the observations in the case $\Omega_{b} h^{2} <0.0223$.

\section{Evolution of vector perturbations in the VT theory}
\label{sec-5} 

In our version of the concordance model 
(GR1 entry in Table \ref{tab:table1}) we have assumed 
$\Omega_{k} = -0.0012 $; nevertheless, this small curvature 
may be neglected. 
It follows from the effects produced by a
much greater curvature with $\Omega_{k} = -0.02$.
These effects (deviations of the dotted line with rerspect to the dotted-dashed one
in the bottom panels of Fig. \ref{figu9}) 
are small and, consequently, the effects of a curvature with 
$\Omega_{k} = -0.0012 $ are very small. On account of
these comments and also for the sake of simplicity,
we have taken $\Omega_{k} = 0 $ all along this section.

Our flat model is chosen to be very similar to the concordance one. 
In order to ensure this similarity, all the parameters are 
assumed to be identical to those of the 
concordance model (see section\ref{sec-3}), excepting 
$\Omega_{\Lambda} $ and $\Omega_{_{A}} $, whose values are assumed to
be $0.7264$ and $0.001$, respectively. This is the configuration
VT16 of Table \ref{tab:table1}. In this case, it may be
easily verified that the relation 
$\Omega_{b} + \Omega_{c} + \Omega_{_{\Lambda}} + \Omega_{_{A}} = 1$ is 
satisfied, as it is mandatory in the flat case.  
Moreover, there is a dominant cosmological constant, and
the dark energy due to $A^{\mu}$ is negligible. 
In spite of this fact, first order perturbations of $A^{\mu}$
might produce important effects. In the absence of these
perturbations the predictions of the configuration VT16 are very similar
to those of the concordance model and, consequently, 
these predictions are compatible with observations.  

We are particularly interested in studying vector perturbations,
which could explain the WMAP anomalies in the first CMB multipoles
according to \cite{mor08}, without alterations of the 
power spectrum $P(k)$ of the energy density perturbations 
(scalar quantity, see \cite{bar80}). A study of these perturbations is
performed in this section. In the linear regime, this study is 
independent of the existence of scalar or tensor perturbations.

In VT theories, 
there are vector perturbations associated to: the peculiar velocity
$v_{i} $, the metric components $h_{i}=g_{0i} $, 
the vector components $A_{i} $,
and the anisotropic stresses $\Pi_{ij}$
\citep{bar80}. As it is usually done, 
these stresses are assumed to be negligible. 

In a flat universe,
vectors $\vec{h} $, $\vec{v} $, and $\vec{A} $ can be expanded in terms of the so-called 
fundamental vector harmonics, whose form is 
$\vec{Q}^{\,\pm} = \vec{\epsilon}^{\,\pm}
\exp (i \vec{k} \cdot \vec{r})$,
where $\vec{k} $ is the wavenumber vector (see \cite{huw97}). 
A representation of vectors 
$\vec{\epsilon}^{\,+}$ and $\vec{\epsilon}^{\,-}$ is \citep{mor07}:
\begin{equation}
\epsilon^{\pm}_{1}=(\pm k_{1} k_{3} /k - i k_{2})/  \sigma \sqrt{2}     \ ,
\end{equation}
\begin{equation}
\epsilon^{\pm}_{2}=(\pm k_{2} k_{3} /k + i k_{1})/ \sigma \sqrt{2}
\end{equation}
\begin{equation}
\epsilon^{\pm}_{3}=\mp  \sigma / k \sqrt{2} \ ,
\end{equation} 
where $\sigma = (k_{1}^{2}+k_{2}^{2})^{1/2}$.
The expansions read as follows $\vec{h} = B^{+} \vec{Q}^{+} + B^{-} \vec{Q}^{-}$, 
$\vec{v} = v^{+} \vec{Q}^{+} + v^{-} \vec{Q}^{-}$, and
$\vec{A} = A^{+} \vec{Q}^{+} + A^{-} \vec{Q}^{-}$. Functions $B^{\pm}(\vec{k},\tau)$,
$v^{\pm}(\vec{k},\tau)$, and $A^{\pm}(\vec{k},\tau)$ describe the perturbation in momentum space. 
The differences
$v_{c}^{\pm}(\vec{k},\tau)=v^{\pm}(\vec{k},\tau)-B^{\pm}(\vec{k},\tau)$ and 
the quantities $A^{\pm}(\vec{k},\tau)$ are gauge invariant. 

In cosmological models based on GR (with and without cosmological constant), 
quantities $A^{\pm}$ vanish and 
the time variations
of $v_{c}^{\pm}$ and $B^{\pm}$ are as follows \citep{mor07}: 
(i) in the matter dominated era, 
\begin{eqnarray}
& &v_{c}^{\pm}(\tau,\vec{k})=v_{c(0)}^{\pm}(\vec{k})/a(\tau) \ , \nonumber \\ 
& & 
B^{\pm}(\tau,\vec{k})=6H_{(0)}^{2} \Omega_{m} v_{c(0)}^{\pm}(\vec{k})/k^{2}a^{2}(\tau)
\label{z}
\end{eqnarray}
and, (ii) in the radiation dominated era,
\begin{eqnarray}
& &v_{c}^{\pm}(\tau,\vec{k})=v_{c}(\tau_{eq},\vec{k})=constant \ , \nonumber \\
& & 
B^{\pm}(\tau,\vec{k})=8 \rho_{r(0)} v_{c}^{\pm}(\tau_{eq},\vec{k})/k^{2}a^{2}(\tau)
\ ,
\label{zz}
\end{eqnarray}
where, $\tau_{eq}$ stands for the conformal time at matter-radiation equivalence.
On account of Eqs.~\ref{z} and \ref{zz}, we conclude that
quantities $v_{c}^{\pm}$ are 
constant (decrease proportional to $a^{-1} $) during the radiation (matter) dominated
era. This is valid for any spatial scale. 
Let us now study the evolution of the gauge invariant quantities 
$v_{c}^{\pm} $ and  $A^{\pm}$ in the VT theory.

The field equations of the theory couple the evolution of $B^{\pm}$, $v^{\pm}_{c}$,
$A^{\pm}$, and the variables $A_{0}$ and $a$ describing the background. 
From the field equations of the VT theory, 
plus the perturbed line element
$ds^{2} = a^{2} ( -d\tau^{2} + \delta_{ij} dx^{i} dx^{j} 
- B^{\pm}Q^{\pm}_{i}dx^{i}d\tau )$,
and the perturbed vector field $A_{\mu} = (A_{0}(\tau ), 
A^{\pm}Q^{\pm}_{i})$, we have found the following equations 
describing --to first order-- the evolution of $v_{c}^{\pm}$, $B^{\pm}$, 
and $A^{\pm}$: 
\begin{equation}
\ddot{A}^{\pm} = k^{2} A_{0} B^{\pm} + \big( 
2 \frac {\ddot{a}}{a} +2 \frac {\dot{a}^{2}}{a^{2}} -k^{2} \big) A^{\pm}
\label{8.1}
\end{equation}  
\begin{eqnarray}
\dot{B}^{\pm} + 2 \frac {\dot{a}}{a} B^{\pm} &=& \frac {32 \pi G \eta}{a^{2}}
\big[ \dot{A}_{0} \big( 2 A_{0} B^{\pm} - A^{\pm} \big) \nonumber \\
& & 
+ A_{0}
\big( A_{0} \dot{B}^{\pm} - \dot{A}^{\pm} \big) \big]
\label{8.2}
\end{eqnarray}  
and
\begin{eqnarray}
B^{\pm} &=& 16 \pi G [ \frac {a^{2}}{k^{2}} \big( \rho_{_{B}} +
p_{_{B}} \big) v_{c}^{\pm} -  \nonumber \\
& & 
\frac {\eta A_{0}}{a^{2}} (
A^{\pm} - A_{0} B^{\pm} ) ]\ .
\label{8.3} 
\end{eqnarray}
From Eqs.~(\ref{7.1}) to~(\ref{7.6}) (after trivial modifications 
necessary to include vacuum energy)
and~(\ref{8.1}) to~(\ref{8.3}) 
one can write a system of first order differential equations 
by using appropriate variables and, then, this system can be 
numerically                            
solved for suitable initial conditions.
These conditions have been fixed in the radiation dominated era, at
redshift $z=10^{8} $. Close to the initial redshift, we have proved that,
approximately, 
all the variables involved in the problem evolve as powers of $\tau $. 
On account of this fact, we have found the growing, constant and decaying modes, and
we have chosen consistent initial conditions.

\begin{figure*}[tb]
\begin{center}
\resizebox{0.8\textwidth}{!}{%
  \includegraphics{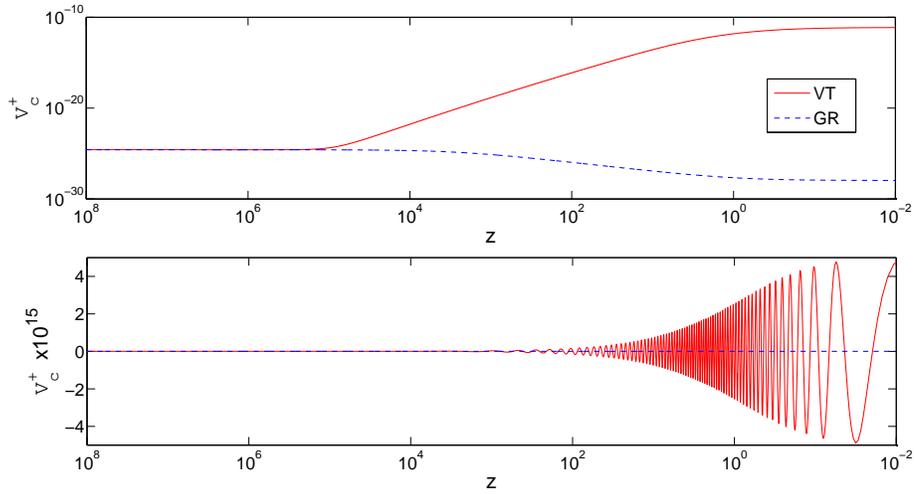}
  }
\caption{Time evolution of the gauge invariant quantity $v_{c}^{+}$
in the VT theory. Top (bottom) panel corresponds to a superhorizon 
(subhorizon) scale whose comoving length is $2\times 10^{4} \ Mpc$
($2\times 10^{2} \ Mpc$). In both panels, the solid (dashed) line 
shows the VT (GR) evolution.}
\label{figu11}       
\end{center}
\end{figure*}

The evolution of quantity $v_{c}^{+}$ is given in Fig. \ref{figu11},
where dashed and solid lines correspond to GR and VT,
respectively. The unique difference between the two panels is  
the spatial scale $L$ ($k=2\pi /L$). In the top panel, this scale is
$L= 2\times 10^{4} \ Mpc$ (superhorizon size). We see that the separation 
between the two lines arises close to the end of the 
radiation dominated era. Moreover, the $v_{c}^{+}$-values reaches the order
$10^{-12} $. After separation, the dashed (solid) line 
displays a decreasing (increasing) $v_{c}^{+}$ quantity. In the bottom panel,
the spatial subhorizon scale is  
$L= 2\times 10^{2} \ Mpc$. In this case, it is evident that the 
solid line (VT theory) oscillates around the dashed line (GR).
We also see that the amplitude of the greatest oscillations is a few times
$10^{-15} $, whereas the order $10^{-12}$ is not reached.
Oscillations (bottom panel) start later than the $v_{c}^{+}$ growing  
associated to the superhorizon scale (top panel).

\begin{figure*}[tb]
\begin{center}
\resizebox{0.8\textwidth}{!}{%
  \includegraphics{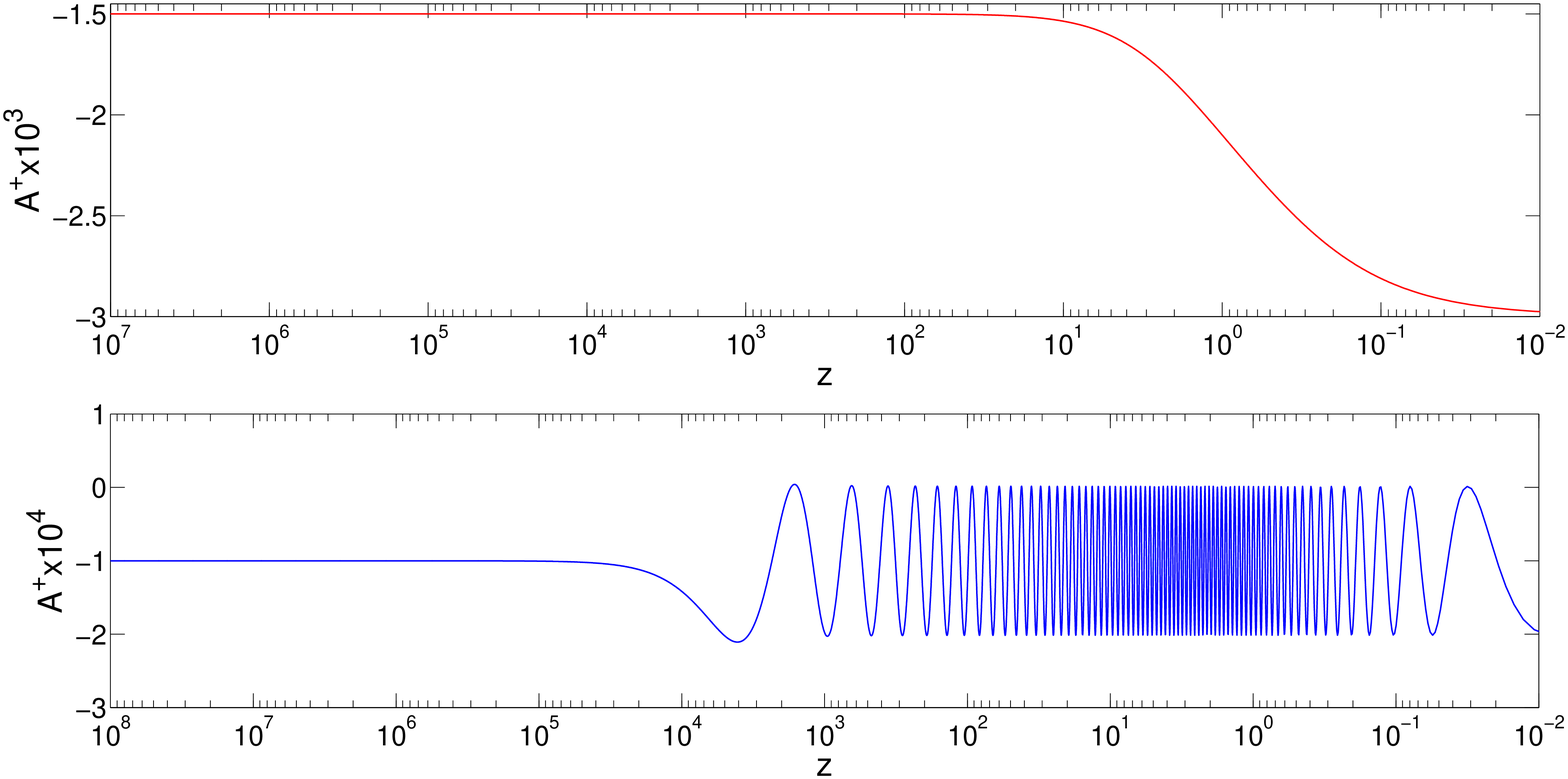}
  }
\caption{Same as in Fig.~\ref{figu11} for the gauge invariant 
quantity $A^{+}$}
\label{figu12}       
\end{center}
\end{figure*}

Fig.~\ref{figu12} \, shows the evolution of $A^{+} $. We see that 
this gauge invariant quantity decreases (oscillates) for 
superhorizon (subhorizon) scales.

\section{Discussion and conclusions}  
\label{sec-6} 

In section \ref{sec-4}, the  VT theory of section \ref{sec-2} has been analyzed 
in detail for negligible $A^{\mu}$-perturbations. Let us first summarize this
analysis. 

There are pairs ($\Omega_{m}$,$h$) explaining the SNe Ia observations;
however, CMB and matter power spectrum observations are not 
simultaneously explained for any of these pairs.

\begin{figure*}[tb]
\begin{center}
\resizebox{0.6\textwidth}{!}{%
  \includegraphics{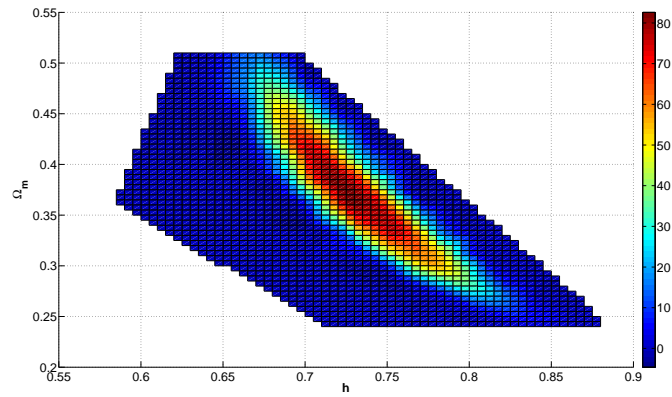}
  }
\caption{Projection of Fig.~\ref{figu5} on the 
($\Omega_{m}$, $h$) plane. The color bar defines the $100p$
values in this 2D representation.}
\label{figu6}       
\end{center}
\end{figure*}

\begin{figure*}[tb]
\begin{center}
\resizebox{0.8\textwidth}{!}{%
  \includegraphics{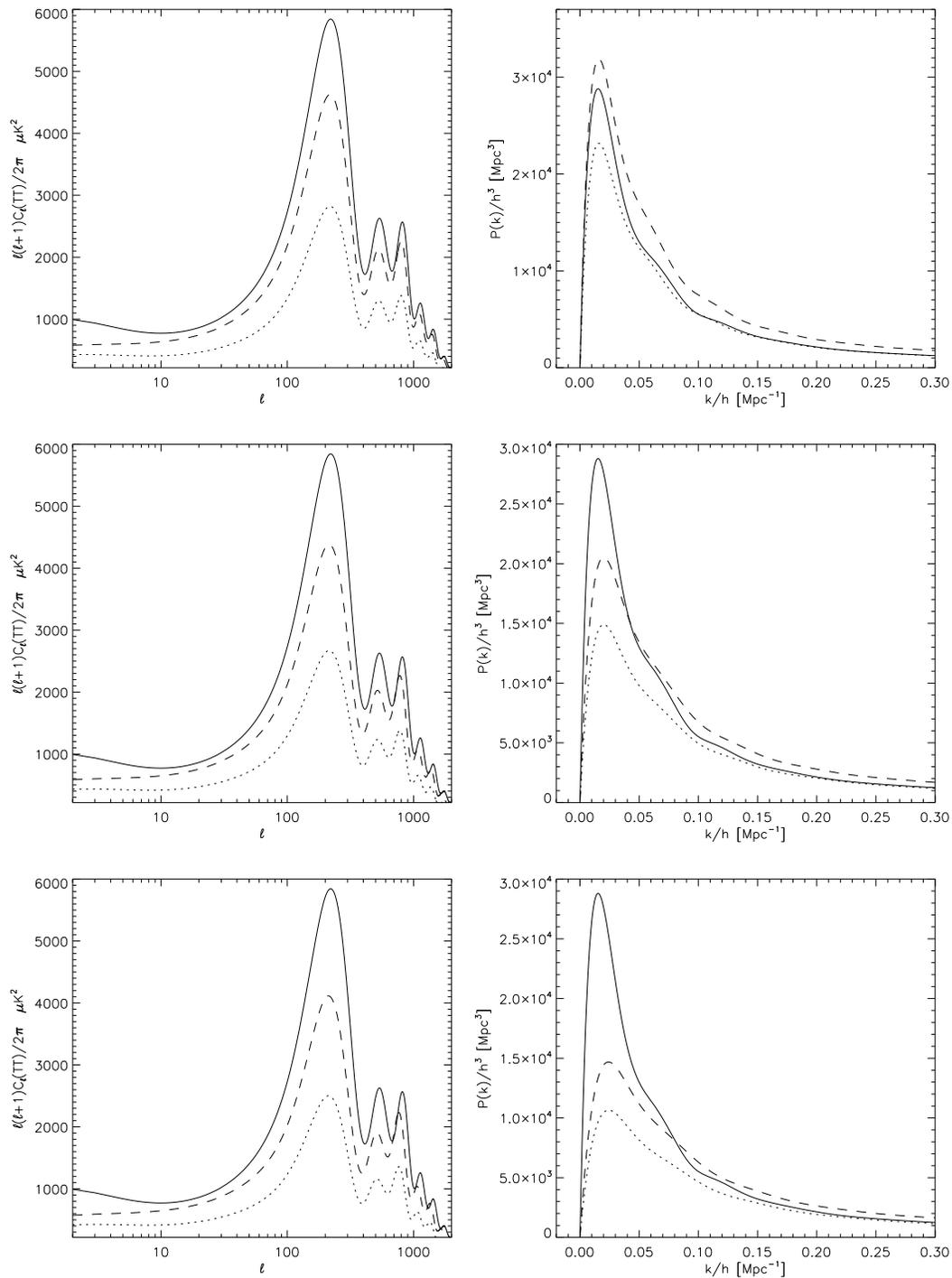}
  } 
\caption{Left: angular power spectrum of the CMB temperature
for some parameter configurations of Table \ref{tab:table1}. Right: power spectrum 
of matter fluctuations for the same cases as in the corresponding 
left panels. The configurations 
considered in each level are:  GR1 (solid), VT1 (dotted),
and VT2 (dashed), in the panels of the top level; 
GR1 (solid), VT10 (dotted), and VT11(dashed) for the middle level; 
and GR1 (solid), VT13 (dotted), and VT14 (dashed) 
for the bottom panels. }
\label{figu8} 
\end{center}      
\end{figure*}

\begin{figure*}[tb]
\begin{center}
\resizebox{0.8\textwidth}{!}{%
  \includegraphics{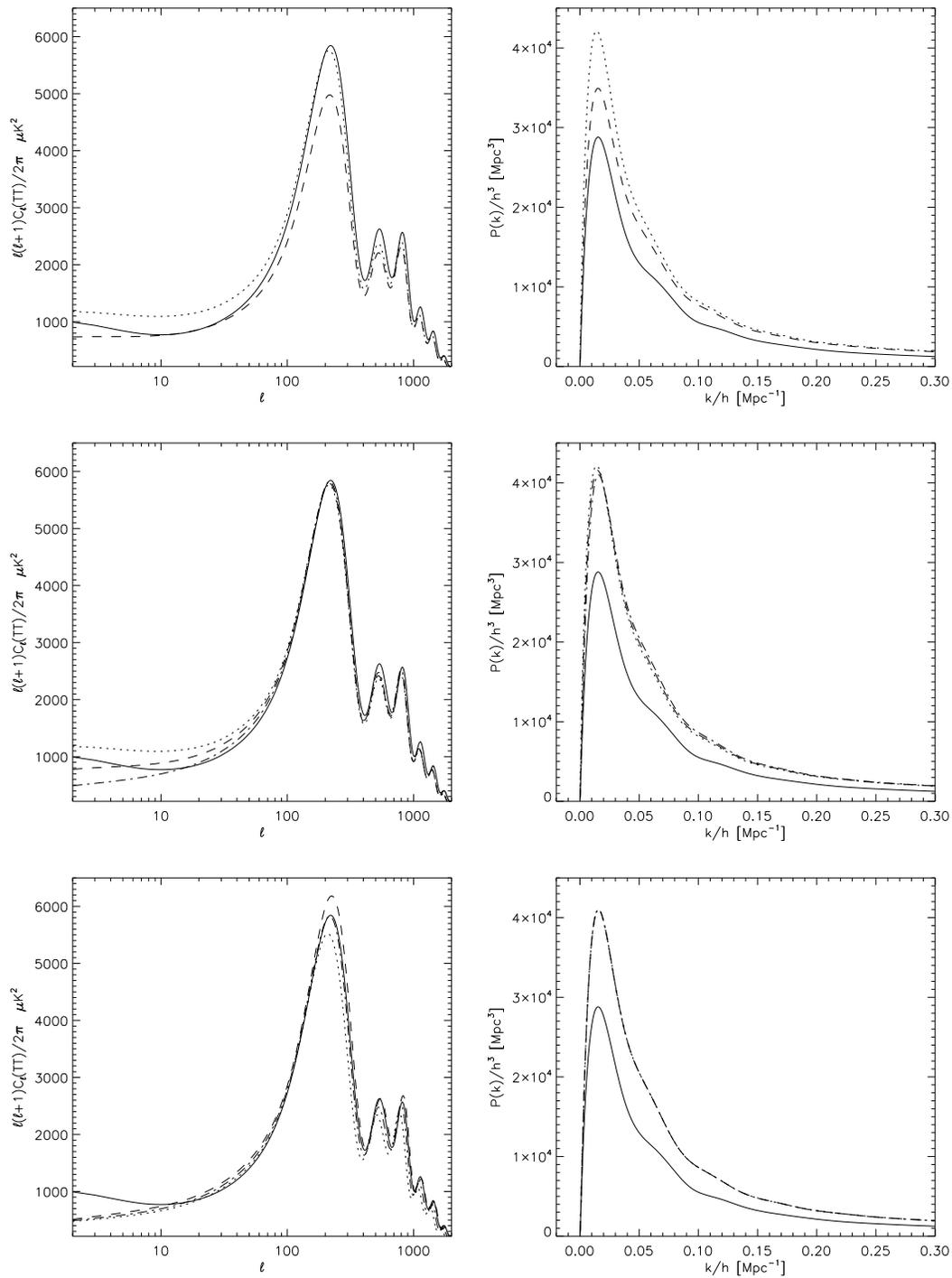}
  }
\caption{Same as in Fig.~\ref{figu8} for the following configurations:
GR1 (solid), VT3 (dashed) and VT4 (dotted) in the top level, 
GR1 (solid), VT4 (dotted), VT5 (dashed), and VT6 (dotted-dashed) 
in the panels of the middle level,
and GR1 (solid), VT6 (dotted-dashed) , VT7 (dotted), and VT8 (dashed) in the bottom panels.
In the bottom right panel cases VT6, VT7 and VT8 are indistinguishable.  }
\label{figu9} 
\end{center}      
\end{figure*}

Although we have studied a set of pairs covering the region of 
the ($\Omega_{m}$,$h$) plane represented in Fig. \ref{figu6},
in which SNe Ia observations are explained with $p>0.05 $,
only the studies corresponding to three of these pairs have been described in detail. 
In these cases, main results are exhibited in Figs. \ref{figu8} 
to \ref{figu10}. 

The main conclusions relative to the CMB angular power spectrum 
are now listed:
(i) there is no a good fitting comparable to that of the concordance model
in the full $\ell$-interval (2,1150), (ii) the best fittings are shown in the dotted lines 
of Fig. \ref{figu10}. These curves correspond to configurations VT9, VT12, and VT15, in which
one has $\sigma_{8} > 0.9$, $\tau =0$, and too large (small) values of $n_{s}$ 
($d n_{s} / d(ln \ k) $); hence, these fittings are found for 
unrealistic values of the cosmological parameters, (iii) even in the 
case $\Omega_{m}=0.25$ and $h=0.82$ (the best fitting), 
the first observed multipoles ($\ell < 10$) are not well explained,
(iv) for other ($\Omega_{m}$,$h$) pairs, the fitting is worse, 
the number of unexplained small $\ell $ multipoles increases and the 
third peak is not well fitted; to see that, the middle left and bottom left panels
of Fig. \ref{figu10} must be compared with the top left one, which corresponds to 
the best fitting, and finally (v) for the parameter configuration VT9
leading to the best fitting to the CMB angular power spectrum, the SNe Ia data
are only marginally explained with $p \sim 0.07$, and the 
Hubble constant is too large ($h=0.82$).
These conclusions strongly suggest that the VT theory under
consideration does not work in the absence of perturbations.  

\begin{figure*}[tb]
\begin{center}
\resizebox{0.6\textwidth}{!}{%
  \includegraphics{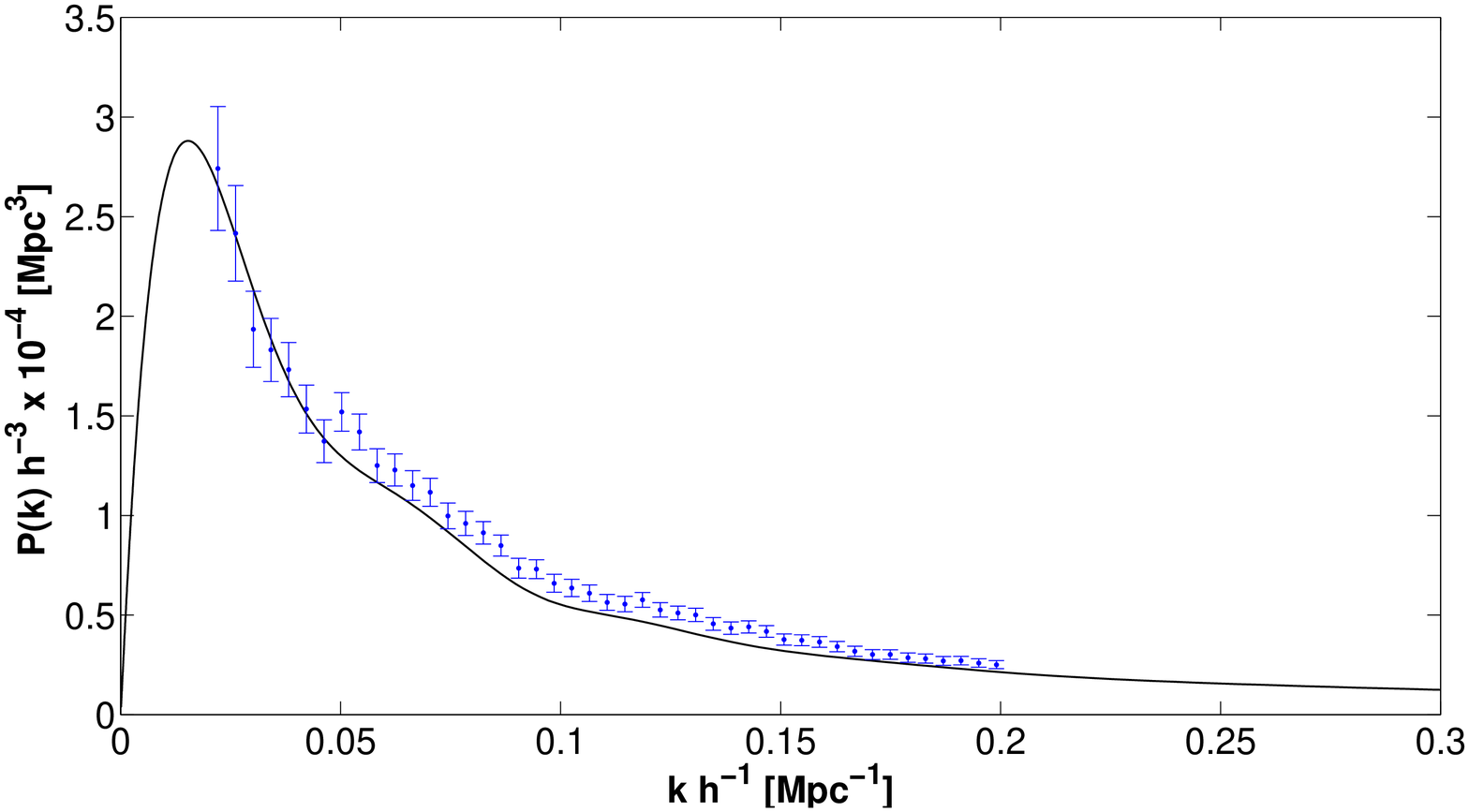}
  } 
\caption{Solid line: all matter power spectrum 
estimated with CAMB for the concordance model. Observational data 
from the SDSS are also shown.
}
\label{figu3}
\end{center}
\end{figure*}

\begin{figure*}[tb]
\begin{center}
\resizebox{0.8\textwidth}{!}{%
  \includegraphics{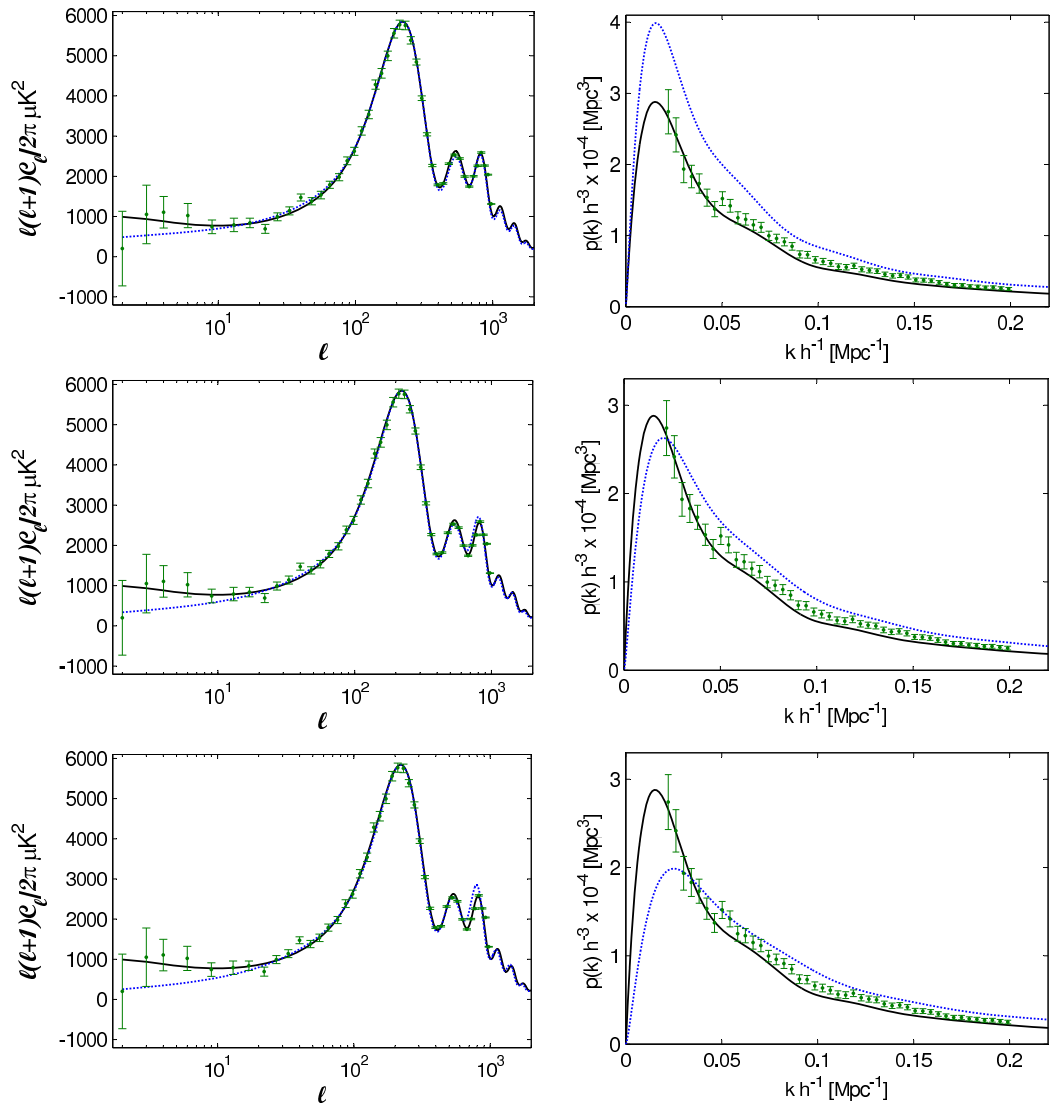}
  }
\caption{Same as in Fig.~\ref{figu8} for the following configurations:
GR1 (solid) and VT9 (dotted) in the top panels, GR1 (solid) and VT12 (dotted) 
in the middle level,
and GR1 (solid) and VT15 (dotted) in the panels of the bottom level. In this figure, 
the same observational data as in Figs.~\ref{figu2} and \ref{figu3}  
have been also displayed in the left and right panels, respectively. }
\label{figu10}       
\end{center}
\end{figure*}

The study of the matter power spectrum also suggest that, in the 
absence of perturbations, the VT 
theory must be rejected. In fact, 
for the best fittings of the CMB angular power spectrum,
the matter power spectrum is not admissible for linear scales 
with $k<0.05h \ Mpc^{-1}$ (see the right panels of Fig. \ref{figu10}); 
however, observations on these scales 
are very well explained by 
the concordance model without any bias between baryonic 
and dark matter (see Fig. \ref{figu3}).

According to the above comments, the VT theory appears to be 
inadmissible for any pair ($\Omega_{m}$,$h$) lying in the region  
represented in Fig. \ref{figu6}, provided that 
the energy of the vector field $A^{\mu}$ plays the role of 
the dark energy, 
there is no either vacuum energy or quintessence, and
$A^{\mu}$-perturbations are negligible.
Typical values of $\Omega_{_{A}} \simeq 1-\Omega_{m}$ may
be seen in Table \ref{tab:table1}. 

A field whose contribution to the background 
energy density of the universe is negligible does not 
significantly affect the background expansion and,
consequently, it does not contribute to the SNe Ia dimming.
However, the fluctuations of this field may produce
crucial effects, for example, anisotropies in the
CMB, which are fully absent in any homogeneous and isotropic 
background. These comments strongly suggest that 
fields being negligible at zero order --in cosmology--
could be fully relevant at higher orders of 
perturbation theory. At first order, there are scalar,
vector, and tensor modes and, if one or more of these 
perturbation components are not negligible, 
the field might be detected by the observation 
of the effects produced by the non-negligible 
modes on the CMB and the matter power spectrum. 
This detection seems to be possible for rather small  
zero order energy densities $\rho_{_{A}}$.
These arguments and those presented in section \ref{sec-1}
--about the possible importance of the VT theories 
in the explanation of the CMB anomalies-- have motivated the study
presented in section \ref{sec-5} on the parameter configuration VT16 of 
Table \ref{tab:table1}, in which  
we have assumed a small background 
energy density with density parameter $\Omega_{_{A}} = 0.001$. Thus,
if both scalar and vector perturbations 
associated to $A^{\mu}$ are negligible, the small value of 
$\Omega_{_{A}} $ --in configuration VT16-- leads to predictions 
in agreement with observations (see section \ref{sec-5}).

After studying the VT theory in the absence of $A^{\mu}$-perturbations,
we have studied the evolution of vector perturbations in 
configuration VT16. This evolution is much
simpler than that of the scalar perturbations and, moreover, both
evolutions are independent (in the linear regime); by this reason,
we have been able to study the evolution of the $A^{\mu}$ vector modes
in section \ref{sec-5}.
The main conclusion is that, for superhorizon scales, quantities 
$v_{c}^{+}$ and $|A^{+}|$ grow. For the scale $L=2\times 10^{4} \ Mpc$, 
these quantities increase 
around $10^{13} $ orders of magnitude from $z=10^{8}$ to present time;
hence, even from small initial superhorizon perturbations, 
we can have large enough perturbations --during the recombination-decoupling 
process-- producing anomalies in the small CMB $\ell $-multipoles.
Subhorizon scales do not grow at the same rhythm and,
furthermore, they oscillate; hence, only the superhorizon 
scales would be relevant at recombination-decoupling
preventing anomalies for too large $\ell$ values.  
The contributions of these superhorizon vector modes to the
low $\ell $ multipoles must be analyzed by using simulations 
and methods similar to those used by \cite{mor08}.

The scalar $A^{\mu}$-perturbations
must be estimated in admissible inflationary models likely triggered 
by appropriate scalar fields coupled or not to field $A^{\mu}$.  
On account of the results obtained in sections \ref{sec-4} and
\ref{sec-5},
any inflationary model leading to negligible scalar $A^{\mu}$-perturbations
must be discarded. The remaining models might be admissible. It 
depends on the effects produced by the 
$A^{\mu}$-perturbations after reheating.
The evolution of the scalar perturbations is to be numerically studied 
with methods similar to those
used in standard cosmology (see \cite{ma95}).

Two are the main conclusions of this paper. The first one is 
important since it rules out, for the first time, 
some versions of the VT theory. In fact,
we have concluded that the VT model studied in 
previous papers \citep{bel08,bel09} --where the  
$A^{\mu}$-energy density plays the role
of the dark energy-- is not compatible with current observations
for negligible $A^{\mu}$-perturbations. This conclusion 
must be taken into account in order to select 
inflationary models in the VT-theory.
Our second conclusion is that model VT16 (see Table \ref{tab:table1})
may explain the presence (absence), at recombination-decoupling, of superhorizon 
(subhorizon) vector modes, which is the key to explain CMB anomalies 
in the low $\ell $ multipoles according to the scheme proposed by 
\cite{mor08}. It has been verified that there are 
growing vector modes in flat configurations different from VT16,
which suggest that the existence of this kind of modes is
a characteristic of the VT theories based on action (\ref{1.1}).
These conclusions motivate 
future investigations in the framework of the VT theory selected in
section \ref{sec-2}, and also in other VT theories.


%
\acknowledgments{This work has been supported by the Spanish
Ministerio de Ciencia e Innovaci\'on, MICINN-FEDER project
FIS2009-07705. We thank J.A. Morales-LLadosa and J. Benavent
for useful discussion.}

\end{document}